\documentstyle[PASJadd,psfig]{PASJ95}
\draft
\begin{document}

\title{ASCA Observations of Three Shakhbazyan's Compact Groups of Galaxies}

\author{
Isao {\sc Takahashi},$^1$
Yasushi {\sc Fukazawa},$^2$
Keiichi {\sc Kodaira},$^3$
Kazuo {\sc Makishima},$^1$\\
Kazuhiro {\sc Nakazawa},$^1$ and Haiguang {\sc Xu}$^4$
\\[12pt]
$^1${\it Department of Physics, School of Science, University of Tokyo,}\\
{\it 7-3-1 Hongo, Bunkyo-ku, Tokyo 113-0033}\\
{\it E-mail(IT): itaka@amalthea.phys.s.u-tokyo.ac.jp}\\
$^2${\it Department of Physics, Faculty of Science, Hiroshima University,}\\
{\it 1-3-1 Kagamiyama, Higashihiroshima, Hiroshima 739-8526}\\
$^3${\it National Astronomical Observatory, Mitaka, Tokyo 181-8588}\\
$^4${\it Department of Applied Physics, Shanghai Jiao Tong University,}\\
{\it 1954 Huashan Road, Shanghai 200030, China}
}

\abst{
X-ray observations of three Shakhbazyan's Compact Groups of Galaxies, SCGG 202, SCGG 205, and SCGG 223, are presented for the first time.
Extended X-ray emission was detected from SCGG 202 and SCGG 223 with 0.5--2 keV luminosity of $1 \times 10^{42}$ erg s$^{-1}$ and $3 \times 10^{42}$ erg s$^{-1}$ (for $H_0$ = 75 km s$^{-1}$ Mpc$^{-1}$), respectively, while no significant emission was detected from SCGG 205 with an upper limit of $0.7 \times 10^{42}$ erg s$^{-1}$.
The X-ray spectra of SCGG 202 and SCGG 223 can be described with thin-thermal plasma emission with temperature about 1 keV.
X-ray properties of SCGG 202 and SCGG 223 are in good agreement with those of other known groups of galaxies, proving the physical nature of their grouping.
}

\kword{Galaxies: cluster of --- X-ray: general --- X-ray: sources}

\maketitle
\thispagestyle{headings}

\section{Introduction} 

Many groups of galaxies have been observed in X-ray so far.
Their X-ray emission has been interpreted to originate from hot intra-group medium (IGM) typically of temperature $\sim$ 1 keV and metal abundance of $\sim$ 0.3 solar (e.g. Mulchaey et al.\ 1993; Fukazawa et al.\ 1998), which is trapped in the gravitational potential binding each group and reflects the group properties.
Concerning compact groups, however, most of targets so far observed in X-rays are so-called Hickson's compact groups of galaxies (HCGs; Hickson 1982). Therefore, our present knowledge of X-ray properties of galaxy groups, especially compact ones,  might be biased by the selection effect.

Shakhbazyan's compact groups of galaxies (SCGGs; Shakhbazyan 1973; Shakhbazyan and Petrosyan 1974; Baier et al.\ 1974; Petrosyan 1974, 1978; Baier and Tiersch 1975, 1976ab, 1978, 1979) comprise a large, relatively homogeneous, catalogue of galaxy groups, almost four times as large as the Hickson's catalogue.
Generally, SCGGs have more member galaxies and a higher fraction of early-type galaxies than HCGs (Tiersch et al.\ 1996).
Therefore, SCGGs are expected to be richer in IGM than HCGs.
Furthermore, it is suggested that galaxies in SCGGs are relatively stronger radio emitters, while those in HCGs are stronger FIR emitters (Tovmassian et al.\ 1999).
Note that only a few SCGGs are included in HCGs.
In spite of these differences, SCGGs have hardly been observed in X-rays so far.
We hence regard SCGGs as valuable targets for X-ray observation for understanding the nature of compact groups of galaxies.

A total of 377 SCGGs were selected from the Palomar Observatory Sky Survey by Shakhbazyan and his colleagues.
The selection criteria of the SCGG catalogue are empirical rather than qualitative, that is;
the group must be compact;
the member galaxies must be compact;
and the group must be isolated.
They typically contain 5-15 member galaxies within a small sky area of less than $10'$ across.
Though they were initially designated as compact groups of compact galaxies (CGCGs), it was later revealed that individual members are not particularly compact (Kodaira et al.\ 1988, 1990; hereafter K90).
In this paper, we therefore use the designation of SCGG. 

Some SCGGs may not be physically bound systems but rather chance projections of galaxies separated along the line of sight.
Furthermore, some members of SCGGs may not be even galaxies, but stellar objects (K90).
In this respect, X-ray observations play an important role, because the X-ray emission directly implies the potential of galaxy groups.

We selected three groups, SCGG 202, SCGG 205, and SCGG 223 for our X-ray study.
SCGG 202 is one of the nearest and the brightest groups among the SCGGs.
The other two targets have been selected based on their relatively near distances and the apparently high velocity dispersions reported in K90.
In this paper, we report on the results of the ASCA observations of these three SCGGs, and compare the obtained properties of them with those of other groups of galaxies.
Optical parameters of these three SCGGs are summarized in table 1 (K90).
In the following, we use the Hubble constant of $H_0 = 75\ h_{75}$ km s$^{-1}$ Mpc$^{-1}$ and $q_0$ = 0.5, and the errors given for X-ray data represent 90\% confidence limits.\\


\section{Observation and Data Reduction} 

We observed SCGG 202, SCGG 205, and SCGG 223 with ASCA (Tanaka et al.\ 1994) during AO-7 phase.
The log of observations is given in table 2.
During all three observations, the GIS (gas imaging spectrometer; Ohashi et al.\ 1996; Makishima et al.\ 1996) were operated in normal PH mode, and the SIS (solid-state imaging spectrometer; Burke et al.\ 1994) were in 1CCD Faint mode. 


For each target, we screened the events under the condition of a telescope viewing direction of $> 5^\circ$ from the dark Earth rim, and a magnetic cutoff rigidity $>$ 8 GeV c$^{-1}$ and $>$ 6 GeV c$^{-1}$ for the GIS and the SIS, respectively.
We further required for the elevation angle from the sunlit Earth rim to be $> 25^\circ$ for the SIS data, in order to avoid the effect of light leakage on the CCD chips.

In figure 1, we show X-ray images taken with the SIS.
Extended emission is clearly detected from SCGG 202 and SCGG 223. 
The peaks of X-ray emission locate at $12^{\rm h}19^{\rm m}53^{\rm s}, 28^\circ 24'59''$ (J2000) and $15^{\rm h}49^{\rm m}47^{\rm s}, 29^\circ 10'33''$ (J2000), respectively, with a typical X-ray positional accuracy of about $0.\hspace{-2pt}'5$.
The X-ray centroid of SCGG 202 locates close to the center of the group, while that of SCGG 223 is at the northeastern region of the group, where six bright member galaxies are concentrated.

In contrast, no significant X-ray emission was detected from SCGG 205.
In fact, we marginally detected a weak X-ray source at $12^{\rm h}35^{\rm m}22^{\rm s}, 27^\circ 35'42''$ (J2000), in the vicinity of galaxy No.6 of SCGG 205 (defined in Baier et al.\ 1975).
This galaxy locates in the western region of the group, where bright member galaxies are found.
However, the source significance is only 3.4 $\sigma$ in terms of the Poisson statistics.
We therefore regard the result on SCGG 205 as a non-detection, and hereafter quote upper limits.


\section{Data Analysis and Results} 

\subsection{Spectral Analysis} 

After the above data reduction, we made spectra for SCGG 202 and SCGG 223.
The events were extracted from circular regions of $6'$ radius from the X-ray centroid.
Concerning SCGG 223, there is a possible contamination source located at $\sim 6'$ to the east in the GIS image, which has not been identified optically.
As the source is outside the SIS chip and just on the GIS grid, we excluded $1'$ around this source for the GIS spectrum.
For background, we used the blank-sky observations, and took the same photon-accumulation region and the same data-selection criteria as were used for the on-source data integration.
After appropriate gain corrections, we added the data from SIS0 and SIS1 into a single SIS spectrum, and those from GIS2 and GIS3 into a single GIS spectrum.
Because of degradation of the SIS, we discarded the SIS data below $\sim$ 0.9 keV.
The obtained GIS and SIS spectra are shown in figure 2, which contain the instrumental response of ASCA.

We jointly fitted the SIS and GIS spectra with a single temperature Raymond-Smith emission model (Raymond \& Smith 1977), modified by photoelectric absorption.
The metal abundances are fixed at 0.3 solar, and the absorption column density $N_{\rm H}$ is fixed at the Galactic value.
For both objects, we obtained acceptable fits.
The results of these fittings are summarized in table 3.



To examine other possibilities, in particular contamination of background active galactic nucleus (AGN), we also fitted the spectra with a power-law model with $N_{\rm H}$ set free.
The results are also shown in table 3.
For SCGG 202, the power-law fit is clearly rejected.
For SCGG 223, though the power-law fit cannot be rejected, the derived photon index $\Gamma$ is too large for an AGN emission.
Fixing $N_{\rm H}$ at the Galactic value hardly affect these results.
Therefore, the Raymond-Smith fit is favored.

For SCGG 205, we only show the upper limits of its flux and luminosity in table 3.
We assumed a typical group emission, represented by the Raymond-Smith model with temperature of 1 keV and metal abundance of 0.3 solar, absorbed with the Galactic column density.

\subsection{Surface Brightness Analysis} 

As the next step, we studied the surface brightness distribution of the X-ray emission from SCGG 202 and SCGG 223, by fitting the radial X-ray profile with $\beta$ model.
In the analysis, we used only the SIS data because of its better spatial resolution.
The $\beta$ model describes projected X-ray surface brightness $S(R)$ as
\begin{equation}
 S(R) = S_0 \left[ 1 + \left(\frac{R}{r_c}\right)^2 \right] ^{-3\beta + 1/2},
\end{equation}
where $S_0$ is the central brightness, $R$ is the projected radius, $r_c$ is the core radius, and $\beta$ is so-called $\beta$ parameter.

In order to properly take into account the instrumental response, we perform a Monte-Carlo simulation (e.g. Kunieda H., Furuzawa A., Watanabe M., the XRT team 1995, http://heasarc.gsfc.nasa.gov/docs/asca/newsletters/xrt3.html) which converts a model source into the simulated ASCA data.
That is, we generate a number of Monte-Carlo photons by specifying their spatial and spectral probability distributions.
The generated photons are then detected by the simulated SIS, according to the known instrumental responses.
This tells us how a ``known'' group looks like when observed with ASCA.

We assume the model group temperature to be spatially constant at the value shown in table 3, and fix its metal abundance at 0.3 solar.
We generated $10^7$ simulated photons, and produced the SIS radial profile exactly in the same way as the actual data.
Then we fitted the actual data with the simulated one, by making only normalization to be a free parameter, and evaluating residuals via the $\chi^2$ tests.
By varying the core radius and $\beta$ parameter of the assumed model, we searched the values which optimized the overall fit goodness.

The fitting results are summarized in figure 3 and table 4.
The sources are significantly extended and the radial profiles of both systems have been fitted well with the $\beta$ model, though we cannot derive the lower limit on the core radius due to insufficient statistics.



\section{Summary and Discussion} 

\subsection{Observed Properties} 

We have detected SCGG 202 and SCGG 223 for the first time in X-rays.
In both objects, we have found that
(1) the X-ray emission center is located within the optical extension of the group,
(2) the emission is extended and the radial surface brightness can be well described by a $\beta$ model,
and (3) the spectrum is well fitted with a single temperature Raymond-Smith model of temperature $\sim$ 1 keV.
We therefore conclude that the observed X-ray emission is attributed to the IGM confined by the group potential, and consequently, that SCGG 202 and SCGG 223 undoubtedly form physically bound systems.

In order to examine their X-ray properties, we plot in figure 4 the IGM temperature versus the X-ray luminosity of the target SCGGs, together with clusters and groups previously observed with ASCA (Fukazawa 1997).
It is clearly seen that SCGG 202 and SCGG 223 obey so-called $kT - L_{\rm x}$ relation of the galaxy clusters and groups.
This implies that the two SCGGs share the same X-ray properties as the other galaxy groups, which in turn can be understood as a smooth extension from those of galaxy clusters.

From SCGG 205, we could not detect significant X-rays.
However this is not puzzling, since a group of galaxies which is slightly less X-ray luminous than SCGG 202 would not have been detected by the present observation at the distance of SCGG 205.
Moreover, some galaxy groups do not show X-ray emissions (Mulchaey et al.\ 1996).


\subsection{Mass Estimation and Comparison with the Velocity Dispersion} 

We can calculate the radially integrated total gravitating mass.
Under the approximation of isothermality, the total mass in the region of three-dimensional radius $r \gg r_c$ is given by (e.g. Sarazin 1988)
\begin{eqnarray}
 M_{\rm tot}(r) &\simeq& \frac{3 \beta k T r}{\mu m_p G} \\
                &  =   & 2.8 \times 10^{13}\ \MO \left(\frac{\mu}{0.6}\right)^{-1}\left(\frac{\beta}{0.5}\right)\left(\frac{k T}{1\ {\rm keV}}\right)\left(\frac{r}{500\ {\rm kpc}}\right),
\end{eqnarray}

where $m_p$ is the proton mass, $k$ is the Boltzmann constant, $G$ is the constant of gravity, and $\mu$ is the mean particle mass in unit of $m_p$.
By taking $\mu = 0.6$ which is a typical value, and using the values of $\beta$ and $kT$ obtained through the fitting (tables 3 and 4), the mass contained in $r < 250\ h_{75}^{-1}\ {\rm kpc}$ becomes $M_{\rm tot} = 0.83 \times 10^{13}\ h_{75}^{-1}\ \MO$ for SCGG 202, and $M_{\rm tot} = 1.6 \times 10^{13}\ h_{75}^{-1}\ \MO$ for SCGG 223.
These values are typical of galaxy groups.
With the total optical luminosity shown in table 1, we can also obtain the mass-to-luminosity ratio as $M/L \sim 103\ h_{75} \MO/\LO$ for SCGG 202 and $M/L \sim 69\ h_{75} \MO/\LO$ for SCGG 223.
These values are again typical of groups of galaxies (e.g. Mulchaey et al.\ 1996).

It has been observed that SCGG 205 and SCGG 223 show rather large radial velocity dispersion, comparable to those of rich clusters (K90).
As a large velocity dispersion means a large virial mass, it would suggest that these SCGGs should have a large amount of dark matter, hence high IGM temperature.
However, their X-ray temperature turned out to be $kT \sim$ 1 keV and gravitating mass-to-luminosity ratio to be $M/L \sim 100\ h_{75} \MO/\LO$, which are comparable to those of typical galaxy groups.
The extremely high velocity dispersion may depend on the group membership, as already pointed out in K90.

For SCGG 202, the velocity dispersion will be reduced from $\sigma_v$ [9] = 456 km s$^{-1}$ to $\sigma_v$ [8] = 221 km s$^{-1}$ if we exclude galaxy No.8 (defined as subgroup SCGG 202A in K90), where [n] means the number of galaxies used for calculation.
The revised velocity dispersion is in good agreement with the plasma temperature determined by X-rays.
This galaxy locates at $\sim 5'$ southwest from the X-ray emission centroid, being one of the farthest galaxies in the projection image of the group.
Therefore, it might not be a proper member of SCGG 202.

In case of SCGG 223, the X-ray centroid locates near the northeast edge of the group, where member galaxies No.1 $\sim$ 6, whose redshift were measured, are concentrated.
If we exclude galaxies No.3 and No.4, whose radial velocities differ from those of the other four by $\sim 2500$ km s$^{-1}$, the velocity dispersion will be reduced from $\sigma_v$ [6] = 1106 km s$^{-1}$ to $\sigma_v$ [4] = 191 km s$^{-1}$.
This suggests that these two galaxies may not be proper members of SCGG 223.\\

In summary, we have observed three SCGGs and detected significant X-ray emission with temperature $\sim$ 1 keV from two of them.
We have confirmed that the main bodies of SCGG 202 and SCGG 223 are not chance projections of galaxies, but really physically bound systems.
Their X-ray properties are similar to those of known other smaller groups and clusters of galaxies.
These results demonstrate the usefulness of X-ray information in understanding the properties of galaxy groups.\\

\clearpage
\section*{Reference}
\re
Baier F.W., Petrosyan M.B., Tiersch H., Shakhbazyan R.K. 1974, Astrofizika 10, 327
\re
Baier F.W., Tiersch H. 1975,  Astrofizika 11, 221
\re
Baier F.W., Tiersch H. 1976a,  Astrofizika 12, 7
\re
Baier F.W., Tiersch H. 1976b,  Astrofizika 12, 409
\re
Baier F.W., Tiersch H. 1978,  Astrofizika 14, 279
\re
Baier F.W., Tiersch H. 1979,  Astrofizika 15, 33
\re
Burke B.E., Mountain R.W., Daniels P.J., Dolat V.S. 1994, IEEE Trans. Nuc. Sci. 41, 375
\re
Fukazawa Y. 1997, Ph.D.thesis, University of Tokyo
\re
Fukazawa Y., Makishima K., Tamura T., Ezawa H., Xu H., Ikebe Y., Kikuchi K., Ohashi T. 1998, PASJ 50, 187
\re
Hickson P. 1982, ApJ 255, 382
\re
Kodaira K., Iye M., Okamura S., Stockton A. 1988, PASJ 40, 533
\re
Kodaira K., Doi M., Ichikawa S., Okamura S. 1990, Publ. Natl. Astron. Obs. Japan 1, 283
\re
Makishima K. Tashiro M., Ebisawa K., Ezawa H., Fukazawa Y., Gunji S., Hirayama M., Idesawa E. et al.\ 1996, PASJ 48, 157
\re
Mulchaey J.S., Davis D.S., Mushotzky R.F., Burstein D. 1993, ApJL 404, 9
\re
Mulchaey J.S., Davis S.D., Mushotzky R.F., Burstein D. 1996, ApJ 456, 80
\re
Ohashi T., Ebisawa K., Fukazawa Y., Hiyoshi K., Horii M., Ikebe Y., Ikeda H., Inoue H. et al.\ 1996, PASJ 48, 157
\re
Petrosyan M.B. 1974, Astrofizika 10, 471
\re
Petrosyan M.B. 1978, Astrofizika 14, 631
\re
Raymond J.C., Smith B.W. 1977, ApJS 35, 419
\re
Sarazin C.L., 1988, X-Ray Emission from Clusters of Galaxies (Cambridge Univ. Press, Cambridge) ch5
\re
Stoll D., Tiersch H., Cordis L. 1997, Astron. Nachr. 318, 7
\re
Stoll D., Tiersch H., Cordis L. 1997, Astron. Nachr. 318, 89
\re
Shakhbazyan R.K. 1973, Astrofizika 9, 495
\re
Shakhbazyan R.K., Petrosyan M.B. 1974, Astrofizika 10, 13
\re 
Tanaka Y., Inoue H., Holt S.S. 1994, PASJ 46, L37
\re
Tiersch H., Stoll D., Neizvestny S., Tovmassian H.M., Navarro S 1996, J. Korean Astron. Soc. 29, 59
\re
Tovmassian H.M., Chavushyan V.H., Verkhodanov O.V., Tiersch H. 1999, ApJ 523, 87

\clearpage

\begin{table*}
\begin{center}
Table~1.\hspace{4pt}Optical parameters of the three SCGGs.
\end{center}
\vspace{6pt}
\begin{tabular*}{\textwidth}{@{\hspace{\tabcolsep}
\extracolsep{\fill}}p{6pc}ccccc}
\hline
\hline\\ [-6pt]
  Object & $\alpha, \delta$ (J2000)$^\ast$ & $n_{gal}^\dagger$ & $z^{\ddagger}$ & $\sigma_v^{\ddagger}$ (km s$^{-1}$) & $L_{\rm tot}^{\S} (h^{-2}_{75} \LO)$ \\[4pt]
\hline\\ [-6pt]
  SCGG 202 \dotfill & No.3 $(12^{\rm h}19^{\rm m}51^{\rm s}\hspace{-4pt}.\hspace{1pt}5, 28^\circ 25'21'')$ & 16 & 0.027 &  456 [9] & $ 8 \times 10^{10}$\\
  SCGG 205 \dotfill & No.6 $(12^{\rm h}35^{\rm m}21^{\rm s}\hspace{-4pt}.\hspace{1pt}6, 27^\circ 35'24'')$ & 14 & 0.096 & 1455 [7] & $45 \times 10^{10}$\\
  SCGG 223 \dotfill & No.2 $(15^{\rm h}49^{\rm m}46^{\rm s}\hspace{-4pt}.\hspace{1pt}5, 29^\circ 10'34'')$ & 13 & 0.083 & 1106 [6] & $23 \times 10^{10}$\\ [4pt]
\hline
\end{tabular*}
\vspace{6pt}\par\noindent
$^\ast$ Positions of the nearest member galaxy from the X-ray peaks (a weak peak for SCGG 205) based on Stoll et al.\ (1997ab).
\par\noindent
$^\dagger$ Number of galaxies constituting the group given in the original catalogue (Baier et al.\ 1975).
These are different from the galaxy members used to calculate $z, \sigma_v,$ and $L_{\rm tot}$.
\par\noindent
$^\ddagger$ Average redshift and radial velocity dispersion in K90, with [$n$] describing the number of galaxies whose redshifts were used for calculation.
\par\noindent
$^{\S}$ Total {\it V}-band luminosity in K90.
\end{table*}

\begin{table*}
\begin{center}
Table~2.\hspace{4pt}Summary of the ASCA observations.
\end{center}
\vspace{6pt}
\begin{tabular*}{\textwidth}{@{\hspace{\tabcolsep}
\extracolsep{\fill}}p{6pc}ccc}
\hline
\hline\\ [-6pt]
  Target & Date & Exposure & Count Rate$^\ast$ \\
   & & (GIS/SIS) & (GIS/SIS) \\[4pt]
\hline\\ [-6pt]
  SCGG 202 \dotfill & 1999 Jun. 7-9 & 58 ks / 44 ks & 9.5 / 24 $^\dagger$\\
  SCGG 205 \dotfill & 1999 Jun. 2-4 & 41 ks / 45 ks & $<$ 1 / $<$ 2 $^\ddagger$ \\ 
  SCGG 223 \dotfill & 1999 Feb. 1-2 & 49 ks / 48 ks & 4.4 / 9.3 $^\dagger$ \\ [4pt]
\hline
\end{tabular*}
\vspace{6pt}\par\noindent
$^\ast$ Background subtracted 0.7-2 keV counts, in $10^{-3}$ c s$^{-1}$ per detector.
\par\noindent
$^\dagger$ Within 6 arcmin of the emission centroid.
Some part of the emission from SCGG 202 may fall outside the SIS chip.
\par\noindent
$^\ddagger$ The integrated region is restricted within 3 arcmin of the weak emission centroid to avoid a contamination source located at about $\sim 5'$ west.
Upper limits corresponding to 99\% confidential levels.
\end{table*}

\begin{table*}
\begin{center}
Table~3.\hspace{4pt}Single-temperature Raymond-Smith$^\ast$ and power-law fits to the joint GIS+SIS spectra.
\end{center}
\vspace{6pt}
\begin{tabular*}{\textwidth}{@{\hspace{\tabcolsep}
\extracolsep{\fill}}lcccrcc}
\hline
\hline\\ [-6pt]
  Object & Model$^\dagger$ & $N_{\rm H}$ & $kT$/$\Gamma^\ddagger$ & $\chi^2/\nu$ & $F_{\rm x}^{\S}$ & $L_{\rm x}^{\S}$ \\
   & & (10$^{20}$ cm$^{-2})$ & (keV)~~~~ & & (erg s$^{-1}$ cm$^{-2}$) & ($h^{-2}_{75}$ erg s$^{-1}$) \\[4pt]
\hline\\ [-6pt]
  SCGG 202 & RS & $1.9 ^{\P}$ & $0.82^{+0.04}_{-0.05}$ & 93/78 & $6.8^{+0.5}_{-0.4} \times 10^{-13}$ & $0.96^{+0.07}_{-0.06} \times 10^{42}$ \\
  & PL & $< 3.4$ & $4.7^{+0.3}_{-0.2}$ & 113/77 & $\cdots$ & $\cdots$ \\
  SCGG 223 & RS & $3.2 ^{\P}$ & $1.14^{+0.12}_{-0.09}$ & 43/45 & $2.3^{+0.3}_{-0.3} \times 10^{-13}$ & $3.1^{+0.4}_{-0.4} \times 10^{42}$ \\
  & PL & $24^{+40}_{-12}$ & $4.3^{+1.3}_{-1.0}$ & 45/44 & $\cdots$ & $\cdots$ \\
\hline\\ [-6pt]
  SCGG 205 & RS & $0.92 ^{\P}$ & 1.0 (fixed) & $\cdots$ &  $< 0.4 \times 10^{-13}$ & $< 0.7 \times 10^{42}$ \\[4pt]
\hline
\end{tabular*}
\vspace{6pt}\par\noindent
$^\ast$ Metal abundance is fixed at 0.3.
Redshift shown in table 1 is taken into consideration.
\par\noindent
$^\dagger$ RS and PL represent Raymond-Smith model and power-law model, respectively.
\par\noindent
$^\ddagger \ kT$ is given for Raymond-Smith model and $\Gamma$ for power-law model.
\par\noindent
$^{\S}$ Fluxes and luminosities in 0.5--2 keV, in terms of the Raymond-Smith model.
\par\noindent
$^{\P}$ Fixed at the Galactic value.
\end{table*}

\begin{table*}
\begin{center}
Table~4.\hspace{4pt}$\beta$ model parameters derived from the surface brightness analysis.
\end{center}
\vspace{6pt}
\begin{tabular*}{\textwidth}{@{\hspace{\tabcolsep}
\extracolsep{\fill}}p{6pc}ccr}
\hline
\hline\\ [-6pt]
  Object & $r_c$ & $\beta$ & $\chi^2/\nu$ \\[4pt]
\hline\\ [-6pt]
  SCGG 202 \dotfill & $0.\hspace{-2pt}'1^{+0.\hspace{-2pt}'4}_{-0.\hspace{-2pt}'1}$ & $0.36^{+0.04}_{-0.01}$ & 20/27 \\
  SCGG 223 \dotfill & $0.\hspace{-2pt}'5^{+1.\hspace{-2pt}'1}_{-0.\hspace{-2pt}'5}$ & $0.50^{+0.25}_{-0.10}$ & 29/27 \\[4pt]
\hline
\end{tabular*}
\end{table*}

\clearpage
\vspace{5cm}
\centerline{Figure Captions}
\bigskip

\begin{fv}{1}{5cm}
{The SIS0+SIS1 images of observed SCGGs in 0.7-2 keV, superposed on the optical images (gray scale).
X-ray images have been smoothed with a Gaussian filter of $\sigma = 0.\hspace{-2pt}'25$.
The contour levels are 1.2, 1.5, 1.8, 2.1, 2.4 and 2.7 for SCGG 202, 0.4 and 0.6 for SCGG 205, and 0.6, 0.9, 1.2 and 1.5 for SCGG 223 in $10^{-3}$ counts s$^{-1}$ arcmin$^{-2}$ unit.
The background is not subtracted, whose level is at $2.7 \times 10^{-4}$ counts s$^{-1}$ arcmin$^{-2}$.
The dashed lines represent the detector edge of the SIS.
$1'$ corresponds to 31 $h_{75}^{-1}$ kpc for SCGG 202, 110 $h_{75}^{-1}$ kpc for SCGG 205, and 97 $h_{75}^{-1}$ kpc for SCGG 223.}
\end{fv}

\begin{fv}{2}{6cm}
{The GIS and SIS spectra of SCGG 202 and SCGG 223, jointly fitted with a single temperature Raymond-Smith model.}
\end{fv}

\begin{fv}{3}{6cm}
{Background-subtracted 0.5--2 keV radial X-ray surface brightness profiles obtained with the SIS.
The instrumental response is not removed.
The dashed histograms represent the best-fit $\beta$ model convolved with the XRT+SIS point spread function.
The solid line shows the simulated point spread function.
$1'$ corresponds to 31 $h_{75}^{-1}$ kpc for SCGG 202, and 97 $h_{75}^{-1}$ kpc for SCGG 223.}
\end{fv}

\begin{fv}{4}{6cm}
{The temperature vs. luminosity diagram of groups and clusters of galaxies taken from Fukazawa (1997).
SCGG 202 and SCGG 223 are shown with circles.
The upper limit for SCGG 205 is also shown with an arrow, for an assumed temperature of 1 keV.}
\end{fv}





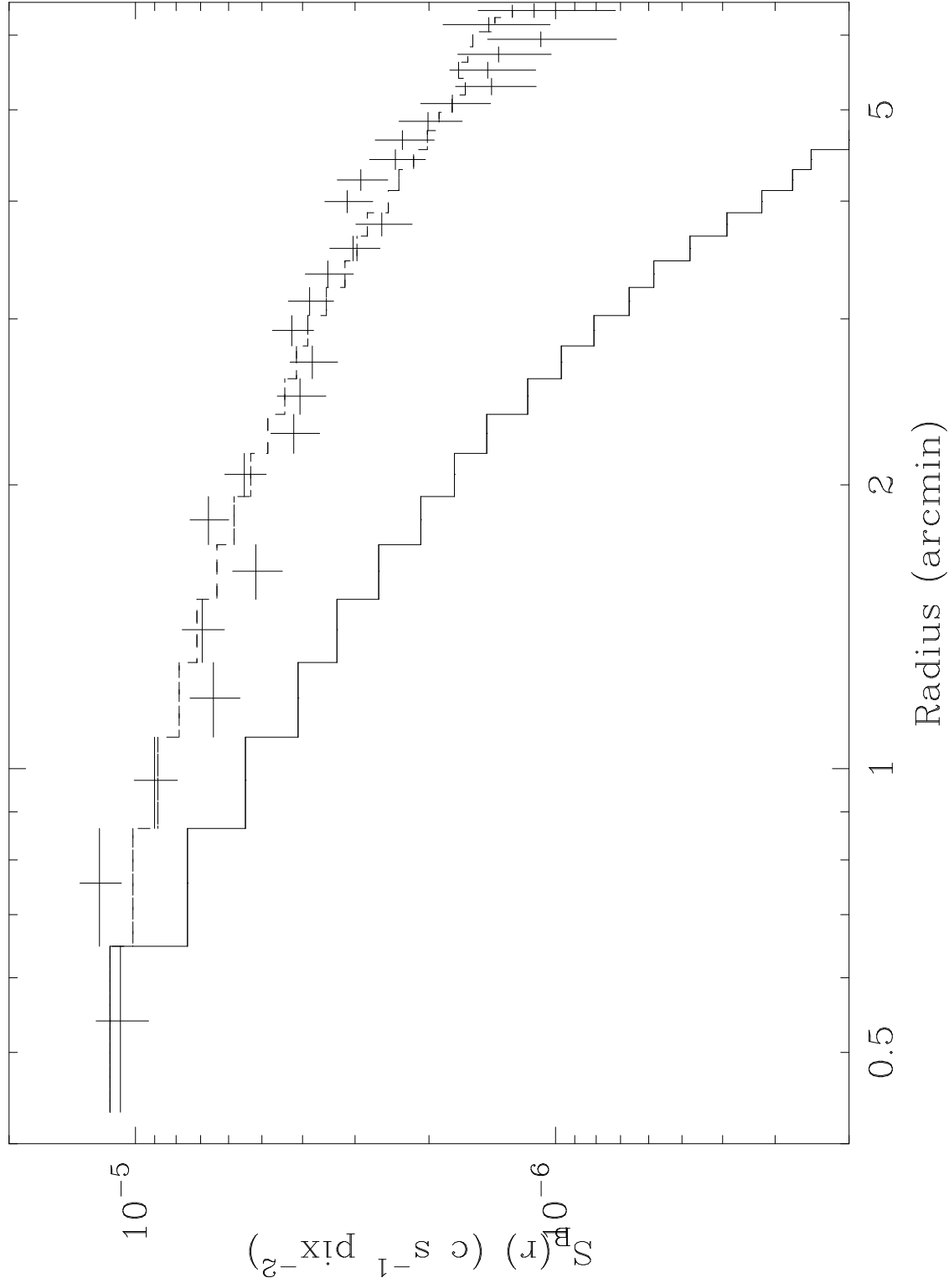
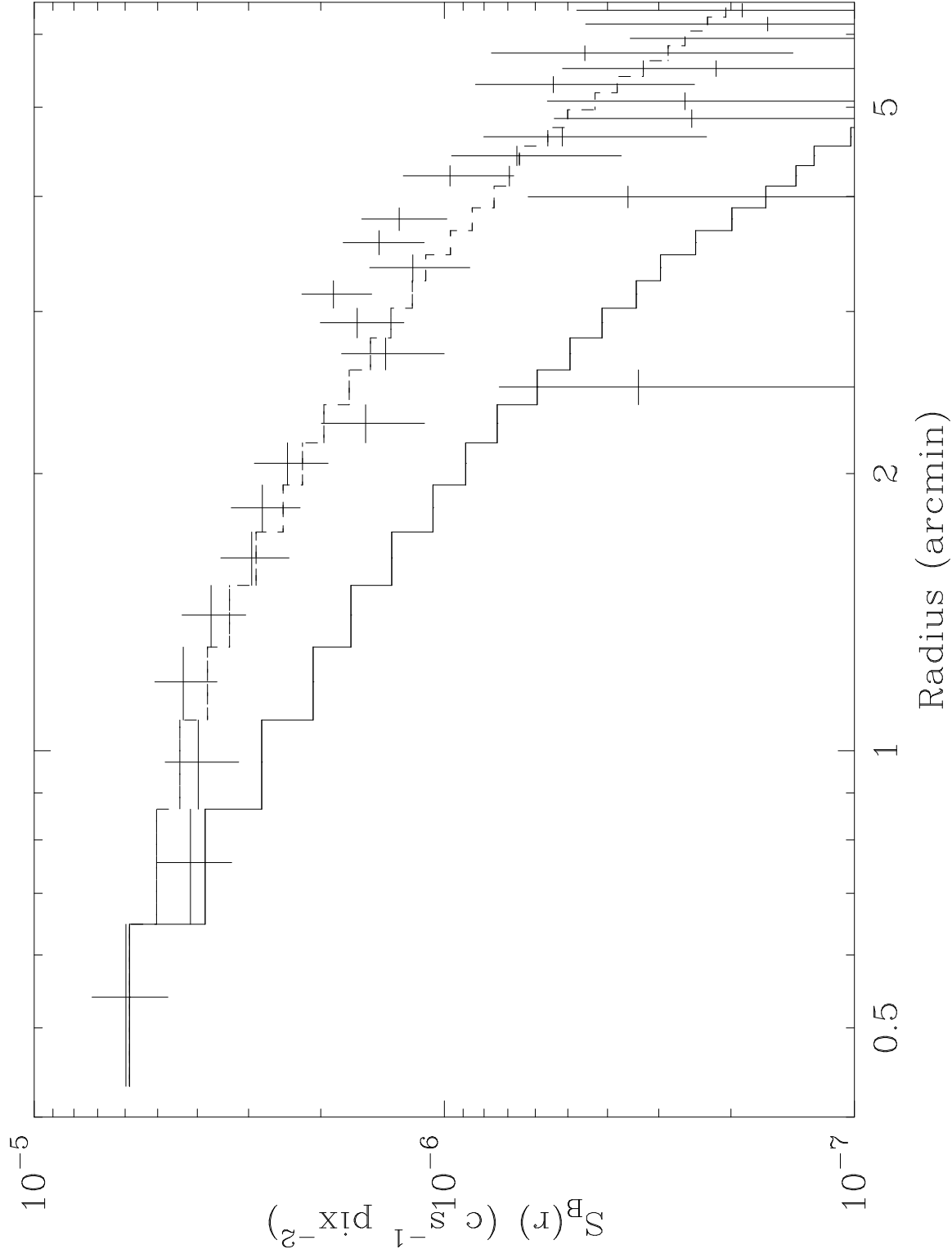
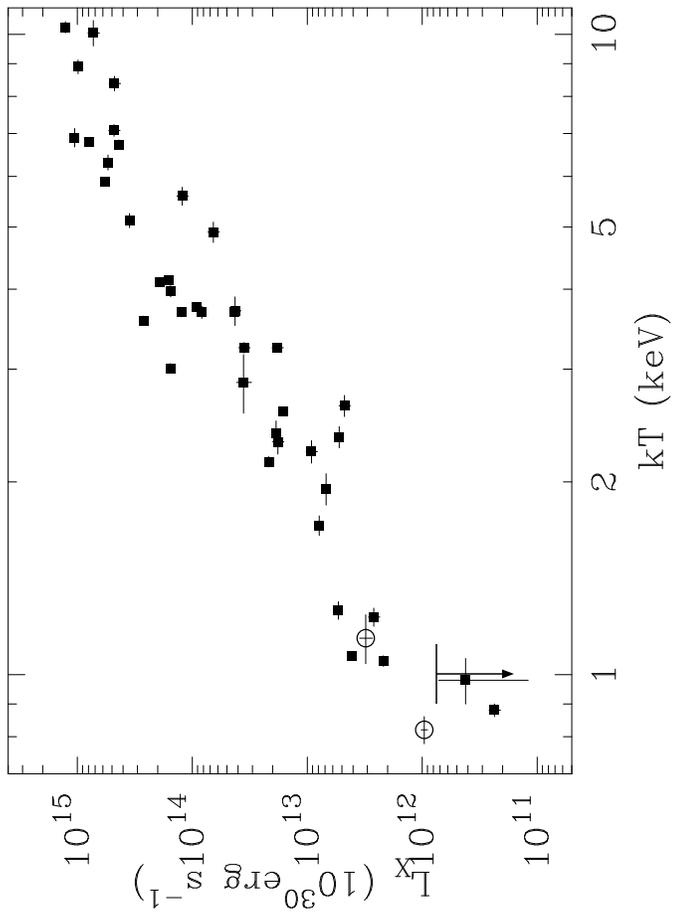
\end{document}